# Enhanced ZnTe infiltration in porous silicon by Isothermal Close Space Sublimation

C. de Melo[a], S. Larramendi[a], V. Torres- Costa[b], J. Santoyo-Salazar[c], M. Behar[d], J. Ferraz Dias[d], O. de Melo[a*]

[a] Physics Faculty, University of Havana, Colina Universitaria, 10400 La Habana, Cuba

[b]Applied Physics Department, Faculty of Sciences, Universidad Autónoma de Madrid, Cantoblanco 28049, Madrid

[c]Physics Department, Centro de Investigación y Estudios Avanzados del Instituto Politécnico Nacional, CINVESTAV-IPN, A.P. 14-740, México D.F. 07360, México

[d]Labortório de Implantação Iônica, Instituto de Fisica, Universidade Federal do Rio Grande do Sul, CP 15051, CEP 91501-970, Porto Alegre, RS, Brazil

Abstract.

Isothermal Close Space Sublimation (ICSS) technique was used for embedding porous silicon (PS) films with ZnTe. It was studied the influence of the preparation conditions and in particular of a chemical etching step before the ZnTe growth, on the composition profile and final porosity of ZnTe embedded PS. The structure of the embedded material was determined by x-ray diffraction analysis while the thickness of the samples was determined by scanning electron microscopy (SEM). Rutherford backscattering (RBS) and Energy Dispersive (EDS) spectrometries allowed determining the composition profiles. We conclude that the etching of the PS surface before the ZnTe growth has two main effects: the increase of the porosity and enhancing the reactivity of the inner surface. It was observed that both effects benefit the filling process of the pores. Since RBS and EDS cannot detect the porosity in the present system, we explore the evolution of porosity by the fitting of the UV-VIS reflectance spectra. The atomic percent determined with this method was in relatively good agreement with that obtained from the RBS and EDS measurements.

[*] Telephone: 00 537 8788950 ext 209/ fax: 00 537 8783471, omelo@fisica.uh.cu

1. Introduction

Porous silicon is a promising matrix host for embedding different materials, such as semiconductors, metals or conductive oxides [1,2] with a wide range of applications. This is because the properties of both the PS and the embedded material can result modified in the final composite. Several techniques have traditionally been used to infiltrate PS films. They include sol- gel [3], multiple step process including impregnation and thermal or chemical treatments [4], atomic layer deposition [5], and electro-deposition [6] among others.

In a previous paper we reported the use of Isothermal Close Space Sublimation (ICSS) technique for embedding PS films with CdSe and ZnTe semiconductors [7]. It was demonstrated that for these semiconductors, approximately constant composition profiles can be obtained inside the pores. In the present paper, we study the influence of the preparation conditions on the compositional profile, morphology, and final porosity of ZnTe embedded PS. X-ray diffraction (XRD), Rutherford backscattering spectrometry (RBS), scanning electron microscope (SEM) and energy dispersive x-rays (EDS) and UV-VIS reflectance spectroscopies were used to characterize the samples.

2. Experimental details

Porous silicon (PS) layers were prepared by electrochemical etching of monocrystalline $p^+$ (100) silicon wafers in a 1:1 ethanol and HF (48% wt.) electrolyte. The process took place in a Teflon cell under illumination provided by a 150 W halogen lamp to increase final porosity. Anodization current was provided by a computer-controlled HG&G 263 galvanostat/ potentiostat. Applied current density was 150 mA/cm$^2$. This set-up is known to produce homogeneous, sponge-like mesoporous silicon samples with pores diameter of a few nanometers [8,9]. The semiconductor ZnTe was grown into the PS matrix by the ICSS technique. General details of this growth setup can be found elsewhere [10]. In short, the growth system is formed by a graphite crucible with two compartments for elemental sources separated by a purge hole. A sliding top part containing the substrate holder allows cyclic exposure of the (1 x 1 cm$^2$) PS layer to Zn (99.99 %) and Te (99.999%) elemental sources (provided by GOODFELLOWS) using a programmed linear actuator (LA12 - PLC LINAK). The growth was performed at 385 $^{o}$C, at a vacuum of $5 \times 10^{-2}$ Pa with 15 cycles. Previously to the growth experiments, some samples were subjected to a chemical etching in a $H_2O$:HF (2:1) solution for 15 s. Exposure time of the substrate to the elemental sources was 30 or 60 s and purge times 5 s.

Rutherford backscattering spectrometry (RBS) analyses were performed at the 3 MV Tandetron accelerator facility of the Ion Implantation Laboratory of the Physics Institute (Federal University of Rio Grande do Sul). For these measurements, 3 MeV α-particles were used. The energy resolution of the Si detectors was 12 keV for alpha particles. The RBS spectra were simulated using the SIMNRA program [11] to determine the compositional profiles.

X- rays diffraction (XRD) patterns in two configuration (Bragg Brentano and grazing incidence) were taken using a Siemens D-5000 powder diffractometer with wavelength (λ) corresponding to Cu $K_{α1}$ radiation. Moreover, Zn and Te compositions within the PS layers were also measured by mapping with a QUANTAX EDS Bruker 127 eV detector, in a field emission gun-scanning electron microscope (FEG-SEM), Carl Zeiss Auriga at 20 kV. RBS and EDS are not able to detect porosity in complex materials as PS. For this reason, Reflectance spectra in the 600- 850 nm range using a Jasco V-560 UV-VIS double-beam spectrophotometer was used to determine the remaining porosity after the ZnTe growth. These measurements were done in an etched sample at three stages: before chemical etching, after chemical etching and after the ZnTe infiltration.

3. Results and discussion

3.1. XRD Experiments

The XRD patterns for a typical ZnTe embedded PS sample is shown in Fig. 1. As it can be seen, the pattern shows diffraction peaks that corresponds to a ZnTe cubic crystalline phase (zinc-blend), evidencing the presence of ZnTe in the PS samples. By using Scherrer equation ( $D = 0.9 \lambda / \beta \cos(\theta)$ ), the size $D$ of the ZnTe crystals can be estimated from the full width at half maximum (β) of the peaks and the corresponding Bragg angle θ. Using the more intense peaks that corresponds to planes with Miller index (111), (220) and (311), we obtained crystallite sizes values of 10.2, 9.0 and 8.4 nm, respectively. The fact that the size of the crystallites is of the same order of the pore diameter of the PS employed seems to indicate that ZnTe is effectively embedded in the mesoporous silicon. To evaluate the amount of ZnTe infiltration, RBS and EDS measurements were performed.

3.2. Rutherford Backscattering Experiments

Typical RBS spectra of the samples are shown in Fig. 2. Both samples A and B were fabricated with 15 cycles and using identical exposure times to the sources (60 s) and purge times (5 s). The sole difference between these samples is that sample B was subjected to a chemical etching step before the growth of ZnTe. To fit the spectra, several layers with different Zn, Te, Si and O composition and thickness were considered. It was assumed that the composite consists of a mixture of Si, $SiO_2$ and ZnTe, i.e. that all oxygen atoms are bound to Si and all Zn and Te atoms are forming ZnTe. This seems to be adequate in view of the stoichiometric 1:1 composition of ZnTe observed along the films in the simulation of RBS spectra (see below). In addition, XRD patterns did not show the presence of any Zn or Te oxide. Then, the profiles were calculated in terms of the composition of the different compounds and not of the single elements. In order to better understand the obtained spectra shown in Fig 2 a) and b), we have also plotted the individual contribution of each element. It can be seen clearly that the Zn/Te intensity relation is 1/3 as expected for stoichiometric ZnTe compound. On the other hand, the simulation allows a good fit of the spectra and demonstrates that ZnTe is distributed along the whole PS layer. In Fig. 3 the compositional profiles are presented. The outermost layer for both samples is composed only by Zn and Te in a stoichiometric ratio of 1:1 although in sample B this layer was thicker than in sample A. In the case of sample A (with PS not chemically etched before infiltration) the concentration of $SiO_2$ varies between 40-80 % while for sample B only between 10-20% and increase toward the PS/substrate interface. This is an expected behavior since the etching step with HF dissolved part of the $SiO_2$. An important result is that in the case of sample A the concentrations of ZnTe inside the PS layer is between 1 and 4 % and for the sample B is much higher ranging from 15 to 80%. Obviously, this fact is attributed to the chemical etching, since ZnTe growth conditions were the same for both samples. We conclude that the etching of the PS layer before the ZnTe growth has two main effects: increase the available porosity and enhance the reactivity of the inner surface. Both effects improve the filling of the pores. ~~Since RBS cannot detect porosity, we explore the evolution of porosity by using reflectance measurements, as it will be shown below.~~

3.3. Electron microscopy and EDS results

Fig. 4 shows a top view SEM image of a pristine PS sample surface. Some pores emerging at the surface are highlighted and their measured diameters are displayed. As can be observed, the pores diameter is about 10 nm. Fig. 5 shows cross-sectional SEM micrographs taken from cleaved ZnTe embedded samples A (a) and B (b). Silicon substrate (bottom) and porous silicon layer are clearly observed together with the ZnTe outermost layer (more perceptible in sample B) confirming the conclusion of RBS analysis. Using EDS, concentrations were measured in the central region of the embedded PS layers. The average concentrations of ZnTe, $SiO_2$ and Si obtained by the RBS spectra simulations and EDS measurements are shown in Table I. A relatively good agreement is observed between the results of both techniques. An EDS compositional map of sample A is shown in Fig. 6 using the characteristic emissions L$\alpha$ for Te and K$\alpha$ for Zn, O and Si respectively. The lower part, where the red colour is predominant, corresponds to the substrate. This map allows visualizing that Zn and Te are effectively distributed along the PS layer in agreement with RBS interpretation.

3.4. Optical reflectance measurements

To complete the study of the infiltration is interesting to know the evolution of the porosity from the pristine PS sample to the ZnTe embedded one. With ICSS the filling of the pores begin in their inner surfaces. For this reason, one would expect that the pores will never be filled completely since complete filling in any region of the pores would obstruct the gas flow along it. However, RBS is not able to detect the porosity because the stopping power (SP) of a void is zero and consequently the total SP of the layer is not affected by the presence of voids. For this reason, we consider using UV- VIS reflectance spectra to estimate the porosity of the PS layer.

It is important to mention here that determining the porosity using only the optical properties is a very difficult task; particularly in a so complex system as the infiltrated PS. In this section we propose a simple model to explain very approximately the evolution of the porosity in each step. In spite of the model simplicity is remarkable that it affords a god fit of the spectra and gives a seemingly satisfactory explanation to the process as it will be shown below.

The refractive index of porous silicon can be approximated by a linear mixing law [12]. Accordingly, if p is the porosity (the volume fraction of air in the sample), $X_{Si}$, $X_{SiO_2}$ and $X_{ZnTe}$ are the volume fractions of Si, SiO$_2$ and ZnTe respectively, then the effective refractive index of the medium is:

$$n_{PS/ZnTe} = p\, n_0 + n_{Si}\, X_{Si} + n_{SiO_2}\, X_{SiO_2} + n_{ZnTe}\, X_{ZnTe} \qquad (1)$$

where $n_0$, $n_{Si}$, $n_{SiO_2}$, $n_{ZnTe}$ are the refractive indexes of air, Si, SiO$_2$ and ZnTe respectively. Expressions for these refractive indexes were obtained by the fit of experimental values [13,14,15] of $n$ vs. $\lambda$ in the range of 600 to 850 nm. As the position of the interference fringes (IF) in the reflectance spectra depends on the refractive index and thickness of the film; a study of these spectra in the different stages of the process is expected to provide information about the filling of the pores, and particularly on the porosity.

Fig. 7 shows reflectance spectra of sample C (grown in the same conditions as sample B but with 30 s of exposure to the sources) in the wavelength range in which IF are observed and in the three stages of the process: i) before the chemical etching; ii) before and iii) after the ZnTe growth. To fit the reflectance spectra of the sample before and after the chemical etching (before ZnTe growth), we consider a partially oxidized PS layer onto a silicon substrate and we used a "single layer" model (multiple waves method) [16] of the reflectance. Reflectance spectra of the sample after the ZnTe growth was not reproduced with this single - layer model because the presence of the outermost thin ZnTe layer. In this case we considered the transfer matrix method [17].

The main information provided by the reflectance spectra is related to the position of the maxima and minima in the IF. It can be observed in Fig. 7 that IF slightly separates after chemical etching, indicating a lower optical density of the layer. It is also observed that after ZnTe growth, IF separation slightly decreases as a consequence of the partial filling of the pores.

Fitting parameters for samples before infiltration were: porosity, thickness and $SiO_2$ volume fractions (VFs) (silicon VF stays determined with the porosity and the $SiO_2$ VF). For samples after ZnTe growth, fit parameters were the porosity of the PS layer and the thickness of both layers, ZnTe infiltrated-PS ($d_1$) and the outer ZnTe layer ($d_2$). It was also considered that the amounts of $SiO_2$ and Si remain without modification after ZnTe growth.

Instead of considering the more complicated task of fitting the absolute value of the reflectance intensity, we look for the coincidence of the wavelength position of the interference maxima and minima between the theoretical model and the experimental spectrum. We assumed that such a procedure would be sufficient for the accurate determination of the parameters of interest in this study. With this aim, the sign function (Sg) of the numerical reflectance increments (ΔR) was built for both the experimental and theoretical spectra. Sg is defined as 1 (ΔR >0), 0 (ΔR=0, IF maxima and minima) or -1 (ΔR <0). Therefore, the fit parameters were calculated by minimizing the square deviation between Sg (ΔR$_{exp}$) of the spectrum and Sg (ΔR$_{sim}$) of the model:

$$\sum_{i=0}^{i_{máx}} \{Sg[\Delta R_{sim}(d_1, d_2, p, X_{SiO_2}, \lambda_i)] - Sg[\Delta R_{exp}(\lambda_i)]\}^2 \quad (2)$$

In the case of the spectra before ZnTe growth, $d_2$ and $X_{ZnTe}$ are obviously zero. After the ZnTe growth $X_{SiO2}$ is a fixed parameter taken from the previous fittings.

The results of the simulations are shown in Fig. 7 by solid lines. In every stage, the calculated values of thickness and porosity are indicated. In the sample embedded with ZnTe, the VF of ZnTe is also specified. It can be noticed that after the chemical etching, the thickness of the PS layer remains almost constant and the porosity increase from 10 to 30 %. This indicates that the main effect of HF was, as expected, to increase the porosity of the PS layer by dissolving part of the $SiO_2$. After the ZnTe growth, the thickness of the ZnTe/PS layer was 1.04 μm (approximately the same than in the previous stage) and that of the ZnTe outermost layer was 164 nm. To fit this spectrum was necessary to consider a porosity of 9 %, substantially smaller than in the previous stage and the VF of ZnTe embedded in PS layer was of 0.21 (volume percent of 21 %). To allow comparison with results of RBS and EDS, this volume percent calculated from the reflectance simulation can be converted to atomic percent using the atomic densities and volume fractions for all the compounds involved: this leads to a value of 42 % in relatively good agreement with RBS and EDS measurements (see Table I or Fig. 3).

4. Discussion and Conclusions

Isothermal Close Space Sublimation (ICSS) technique was used for embedding PS films with ZnTe. The films were characterized by different techniques as x-ray diffraction (XRD), Rutherford backscattering spectrometry (RBS), energy dispersive x-ray (EDS) and optical reflectance spectroscopies. These techniques allowed us to describe the main aspects of the filling process of the pores as discussed below.

During the exposure to one of the two elemental sources, Zn or Te vapors infiltrate along the pores of PS. Since in ICSS the temperature of the sources is the same as that of the substrate, the deposition is self-regulated and it is expected that only a very thin film will be adsorbed in the inner surfaces of the pores. Then, the filling process would progress by engrossing this very thin film in successive deposition cycles; thus, the internal surface of the pores should be covered in a conformal way. Then, a certain degree of porosity must remain in the embedded samples. The above assumptions were demonstrated by the analysis of the evolution of the porosity in the different stages of the process. Then, ZnTe fills up only part of the void volume. Deposition also occurs in the external surface of the sample where, as was verified, an outermost film of pure ZnTe forms.

It is expected that the etching enlarge the pores of the sample, facilitating the flow of gas inside the pores, because HF etching dissolves part of the SiO2 on the inner pore walls. This was verified by reflectance measurements in PS samples before and after chemical etching. But the dissolution of in-pore $SiO_2$ alone can't account for the increased amount of material embedded into PS. In fact, by enlarging the pore size, the number of pores can actually decrease due to the complete dissolution of the thinnest pore walls. This would lead to an actual reduction of the PS internal surface available for adsorption and of the amount of adsorbed material. Then we conclude that the role of HF etching, in relation to the enhanced infiltration, is rather associated with an increase of the inner surface reactivity favoring the adsorption of ZnTe. This is a consequence of SiO2 partial removal in the inner surfaces in which fresh more reactive silicon sites become available. In fact, in the case of samples etched before ZnTe growth, both the outermost layer of ZnTe and the amount of ZnTe embedded in the porous increase significantly. It is worth to note that this growth procedure allows obtaining in a simple way a Si/PS-ZnTe/ZnTe heterostructure which could be useful for devices fabrication.

Acknowledgements. The authors acknowledge the technical assistance of S. de Roux in the preparation of ZnTe embedded samples. This work was to a great extent supported by the CAPES-MES project No. 220/12. OdM and VTC acknowledge the support given by the agreement between the University of Havana and the Universidad Autónoma de Madrid.

Table I. ZnTe, SiO$_2$ and Si atomic percent for samples A and B determined by EDS and RBS.

|  | Sample A | | Sample B | |
| --- | --- | --- | --- | --- |
|  | **EDS** | **RBS** | **EDS** | **RBS** |
| *ZnTe (at. %)* | 5.9 ± 0.6 | 2.5 ± 0.5 | 47.3 ± 2.3 | 45.1 ± 0.1 |
| *SiO$_2$ (at. %)* | 50.2 ± 4.8 | 52.9 ± 1.4 | 16.0 ± 1.9 | 18.8 ± 0.9 |
| *Si (at. %)* | 43.9 ± 5.1 | 54.7 ± 3.3 | 36.7 ± 3.0 | 36.1 ± 1.5 |

Figure captions.

Fig. 1. X-ray diffractogram for a typical ZnTe embedded PS sample showing the characteristic peaks of Zinc Blend ZnTe.

Fig. 2. Rutherford backscattering spectra (dot black curves) and simulation by SIMNRA software (solid red curve) of sample A (non-chemically etched before the growth) and sample B (chemically etched before the growth). Spectra corresponding to individual elements are also presented. In the inset of a), an amplified view of the region of Zn and Te spectra is shown.

Fig. 3. Compositional profiles of samples A and B obtained by the simulation of RBS spectra.

Fig. 4. Top view SEM image of a pristine PS sample surface showing the measured diameter of the emerging pores.

Fig. 5. Cross section SEM micrograph for: a) sample A and b) sample B. The thickness of the ZnTe/PS layer is indicated.

Fig. 6. EDS compositional map of sample A using characteristic x- ray emissions (L$\alpha$ for Te and K$\alpha$ for Zn, O and Si, respectively)

Fig. 7. Reflectance spectra of the sample: before chemical etching (upper), after chemical etching (middle) and after ZnTe growth (lower). Solid curves represent the simulated reflectance spectra. $X_{ZnTe}$ represents the volume fraction of ZnTe expressed in percent; it is equivalent to an atomic % of 42 %.

Appendix.

1. Multiple waves – single layer model

The expression for the optical reflectance considers the interference of the multiple rays that come out from the front surface of the sample, considering normal incidence of the light. Two interfaces were considered: air-PS and PS-silicon. The reflection coefficient for multiple reflections can be expressed as:

$$r = \frac{r_1 + r_2 e^{-i\delta}}{1 + r_1 r_2 e^{-i\delta}} = \frac{r_1 + r_2 e^{-i\,4\pi\, n_1\, d/\lambda}}{1 + r_1 r_2 e^{-i\,4\pi\, n_1\, d/\lambda}} \qquad (1)$$

And the reflectance R is given by:

$$R = r\,\tilde{r} = \frac{r_1^2 + r_2^2 + 2\, r_1\, r_2 \cos\delta}{1 + r_1^2 r_2^2 + 2\, r_1\, r_2 \cos\delta} \qquad (2)$$

In these equations $r_1$ and $r_2$ are the reflection coefficient corresponding to the interfaces air- PS and PS- silicon; $\delta = 4\pi\, n_1 d/\lambda$ is the phase difference introduced when it crosses the film in the two directions; $n_1$ is the refraction index of the porous silicon and d is the thickness of the layer.

2. Two - layer model. Transference matrix method

Three interfaces were considered: air-ZnTe (I), ZnTe-PS (II) and SP-silicon (III). The boundary conditions of the electric $(\vec{E})$ and magnetic $(\vec{H})$ fields establish that their tangential components have to be continuous through the boundaries. Expressing these conditions in matrix form is possible to obtain the relation between the tangential components of the field in the frontier I and frontier III:

$$\begin{bmatrix} E_I \\ H_I \end{bmatrix} = \mathbb{M} \begin{bmatrix} E_{III} \\ H_{III} \end{bmatrix} \qquad (3)$$

The matrix ($\mathbb{M}$) elements are:

$$m_{11} = Cos[\delta_1]\, Cos[\delta_2] - Sin[\delta_1]\, Sin[\delta_2](Y_2/Y_1)$$

$$m_{12} = i(Cos[\delta_2]Sin[\delta_1]/Y_1 + Cos[\delta_1]Sin[\delta_1]/Y_2)$$

$$m_{21} = i(Cos[\delta_2]Sin[\delta_1]Y_1 + Cos[\delta_1]Sin[\delta_1]Y_2)$$

$$m_{22} = Cos[\delta_1]Cos[\delta_2] - Sin[\delta_1]Sin[\delta_2](Y_1/Y_2)$$

Where: $\delta_1 = 2\pi n_1 d_1/\lambda$, $\delta_2 = 2\pi n_2 d_2/\lambda$, $Y_1 = \sqrt{\varepsilon_0/\mu_0}\, n_1$ y $Y_2 = \sqrt{\varepsilon_0/\mu_0}\, n_2$.

$n_1, n_2, d_1$ and $d_2$ are the refractive index and the thickness of the ZnTe and PS layers, respectively. $\varepsilon_0$ and $\mu_0$ are the permittivity and the permeability of vacuum.

Taking into account this expression is possible obtain the reflection coefficient:

$$r = \frac{Y_0 m_{11} + Y_0 Y_s m_{12} - m_{21} - Y_s m_{22}}{Y_0 m_{11} + Y_0 Y_s m_{12} + m_{21} + Y_s m_{22}} \quad (4)$$

Here $Y_0 = \sqrt{\varepsilon_0/\mu_0}\, n_0$ and $Y_s = \sqrt{\varepsilon_0/\mu_0}\, n_s$ while $n_0$ and $n_s$ are the refractive index of the air and the silicon substrate, respectively. Then the reflectivity is calculated as $r\tilde{r}$.

RBS cannot detect porosity, we explore the evolution of porosity by using reflectance measurements, as it will be showed below.

3.3. Electron microscopy and EDS results

Fig. 4 shows cross-sectional SEM micrographs of samples A (a) and B (b) taken from cleaved samples. Silicon substrate (bottom) and porous silicon layer are clearly observed together with the ZnTe outermost layer (more perceptible in sample B) confirming the conclusion of RBS analysis. Using EDS, concentrations were measured in the central region of the embedded PS layers. The average concentrations of ZnTe, SiO$_2$ and Si obtained by the RBS spectra simulations and EDS measurements are shown in Table I. A relatively good agreement is observed between the results of both techniques. An EDS compositional map of sample A is shown in Fig. 5 using the characteristic emissions L$\alpha$ for Te and K$\alpha$ for Zn, O and Si respectively. The lower part, where the red colour is predominant, corresponds to the substrate. This map allows visualizing that Zn and Te are effectively distributed along the PS layer in agreement with RBS interpretation.

3.4. Optical reflectance measurements

To complete the study of the infiltration is interesting to know the evolution of the porosity from the pristine PS sample to the ZnTe embedded one. With ICSS the filling of the pores begin in their inner surfaces. For this reason one would expect that the pores will never be filled completely since complete filling in any region of the pores would obstruct the gas flow along it. However, RBS is not able to detect the porosity because the stopping power (SP) of a void is zero and consequently the total SP of the layer is not affected by the presence of voids. For this reason, we consider using UV- VIS reflectance spectra to determine the porosity of the PS layer. The refractive index of porous silicon can be approximated by a linear mixing law [11]. Accordingly, if p is the porosity (the volume fraction of air in the sample), $X_{Si}$, $X_{SiO_2}$ and $X_{ZnTe}$ are the volume fractions of Si, SiO$_2$ and ZnTe respectively, then the effective refractive index of the medium is:

$$n_{PS/ZnTe} = p\, n_0 + n_{Si} X_{Si} + n_{SiO_2} X_{SiO_2} + n_{ZnTe} X_{ZnTe} \tag{1}$$

where $n_0$, $n_{Si}$, $n_{SiO_2}$, $n_{ZnTe}$ are the refractive indexes of air, Si, SiO$_2$ and ZnTe respectively. Expressions for these refractive indexes were obtained by the fit of experimental values [12,13,14] of n vs. $\lambda$ in the range of 600 to 850 nm. As the position of the interference fringes (IF) in the reflectance spectra depends on the refractive index and thickness of the film; a study of these spectra in the different stages of the process is expected to provide information about the filling of the pores, and particularly on the porosity.

Fig. 6 shows reflectance spectra of sample C (grown in the same conditions as sample B but with 30 s of exposure to the sources) in the wavelength range in which IF are observed and in the three stages of the process: i) before the chemical etching; ii) before and iii) after the ZnTe growth. To fit the reflectance spectra of the sample before and after the chemical etching (before ZnTe growth), we consider a partially oxidized PS layer onto a silicon substrate and we used a "single layer" model (multiple waves method) [15] of the reflectance. Reflectance spectra of the sample after the ZnTe growth was not reproduced with this single - layer model because the presence of the outermost thin ZnTe layer. In this case we considered the transfer matrix method [16].

The main information provided by the reflectance spectra is related to the position of the maxima and minima in the IF. It can be observed in Fig. 6 that interference fringe slightly separates after chemical etching, indicating a lower optical density of the layer. It is also observed that after ZnTe growth, IF separation slightly decreases as a consequence of the partial filling of the pores.

Fitting parameters for samples before infiltration were: porosity, thickness and Si and SiO$_2$ volume fractions. For samples after ZnTe growth, fit parameters were the porosity of the PS layer and the thickness of both layers, ZnTe infiltrated-PS ($d_1$) and the outer ZnTe layer ($d_2$). It was also considered that the amount of SiO$_2$ remains without modification after ZnTe growth.

Instead of considering the more complicated task of fitting the absolute value of the reflectance intensity, we look for the coincidence of the wavelength position of the interference maxima and minima between the theoretical model and the experimental spectrum. We assumed that such a procedure would be sufficient for the accurate determination of the parameters of interest in this study. With this aim, the sign function (Sg) of the numerical reflectance increments ($\Delta R$) was built for both the experimental and theoretical spectra. Sg is defined as 1 ($\Delta R > 0$), 0 ($\Delta R = 0$, IF maxima and minima) or -1 ($\Delta R < 0$). Therefore, the fit parameters were calculated by minimizing the square deviation between Sg ($\Delta R_{exp}$) of the spectrum and Sg ($\Delta R_{sim}$) of the model:

$$\sum_{i=0}^{i_{máx}} \{Sg[\Delta R_{sim}(d_1, d_2, p, X_{SiO_2}, \lambda_i)] - Sg[\Delta R_{exp}(\lambda_i)]\}^2 \qquad (2)$$

In the case of the spectra before ZnTe growth, $d_2$ is obviously zero.

The results of the simulations are shown in Fig. 6 by color lines. In every stage, the calculated values of thickness and porosity are indicated. In the sample embedded with ZnTe, the volume fraction of ZnTe is also specified. It can be noticed that after the chemical etching, the thickness of the PS layer remains almost constant and the porosity increase from 10 to 30 %. This indicates that the main effect of HF was, as expected, to increase the porosity of the PS layer by dissolving part of the $SiO_2$. After the ZnTe growth, the thickness of the ZnTe/PS layer was 1.04 µm (approximately the same than in the previous stage) and that of the ZnTe outermost layer was 164 nm. To fit this spectrum was necessary to consider a porosity of 9 %, substantially smaller than in the previous stage and the volume fraction of ZnTe embedded in PS layer was of 21 %. To allow comparison with results of RBS and EDS, this volume percent calculated from the reflectance simulation can be converted to atomic percent using the atomic densities and volume fractions for all the compounds involved: this leads to a value of 42 % in relatively good agreement with RBS and EDS measurements (see Table I or Fig. 3).

4. Discussion and Conclusions

Isothermal Close Space Sublimation (ICSS) technique was used for embedding PS films with ZnTe. The films were characterized by different techniques as x-ray diffraction (XRD), Rutherford backscattering (RBS), Energy Dispersive X-ray (EDS) and optical reflectance spectrometries. These techniques allowed us to describe the main aspects of the filling process of the pores as discussed below.

During the exposure to one of the two elemental sources, Zn or Te vapors infiltrate along the pores of PS. Since in ICSS the temperature of the sources is the same as that of the substrates, the deposition is self-regulated and only a very thin film is adsorbed in the inner surfaces of the pores. Then, the filling process progresses by engrossing this very thin film in successive deposition cycles and, in this way, the internal surface of the pores is covered in a conformal way. For this reason, a certain degree of porosity remains in the embedded samples, as it was demonstrated by the analysis of the evolution of the porosity in the different stages of the process. Then, ZnTe fills up only part of the void volume. Deposition also occurs in the external surface of the sample where, as was verified, an outermost film of pure ZnTe forms.

It is expected that the etching increases the porosity of the sample, facilitating the flow of gas inside the pores, because HF etching dissolves part of the $SiO_2$. This was verified by reflectance measurements in PS samples before and after chemical etching that shows an increment of the porosity in the etched sample. But only this effect could not explain the strong influence of the etching in the filling process. In fact, in the case of samples etched before ZnTe growth, both the outermost layer of ZnTe and the amount of ZnTe embedded in the porous increase significantly. We concluded that the HF etching also increase the reactivity of the surface favoring the surface adsorption of ZnTe and then the filling of the pores. It is worth to note that this growth procedure allows obtaining in a simple way a Si/PS-ZnTe/ZnTe heterostructure which could be useful for devices fabrication.

Acknowledgements. The authors acknowledge the technical assistance of S. de Roux in the preparation of ZnTe embedded samples. This work was to a great extent supported by the CAPES-MES project No. 220/12. OdM and VTC acknowledge the support given by the agreement between the University of Havana and the Universidad Autónoma de Madrid.

Table I. ZnTe, SiO$_2$ and Si molar concentrations for samples A and B determined by EDS and RBS.

|  | Sample A | | Sample B | |
| --- | --- | --- | --- | --- |
|  | EDS | RBS | EDS | RBS |
| *ZnTe (at. %)* | 5.9 ± 0.6 | 2.5 ± 0.5 | 47.3 ± 2.3 | 45.1 ± 0.1 |
| *SiO$_2$ (at. %)* | 50.2 ± 4.8 | 52.9 ± 1.4 | 16.0 ± 1.9 | 18.8 ± 0.9 |
| *Si (at. %)* | 43.9 ± 5.1 | 54.7 ± 3.3 | 36.7 ± 3.0 | 36.1 ± 1.5 |

Figure captions.

Fig. 1. X-ray diffractogram for a typical ZnTe embedded PS sample showing the characteristic peaks of Zinc Blend ZnTe.

Fig. 2. Rutherford backscattering spectra (dot black curves) and simulation by SIMNRA software (solid red curve) of sample A (non-chemically etched before the growth) and sample B (chemically etched before the growth). Spectra corresponding to individual elements are also presented. In the inset of a), an amplified view of the region of Zn and Te spectra is shown.

Fig. 3. Compositional profiles of samples A and B obtained by the simulation of RBS spectra.

Fig. 4. Cross section SEM micrograph for: a) sample A and b) sample B. The thickness of the ZnTe/PS layer is indicated.

Fig. 5. EDS compositional map of sample A using characteristic x-ray emissions (L$\alpha$ for Te and K$\alpha$ for Zn, O and Si, respectively)

Fig. 6. Reflectance spectra of the sample: before chemical etching (upper), after chemical etching (middle) and after ZnTe growth (lower). Solid curves represent the simulated reflectance spectra. $X_{ZnTe}$ represents the volume fraction of ZnTe expressed in percent; it is equivalent to an atomic % of 42 %.

Appendix.

1. Multiple waves – single layer model

The expression for the optical reflectance considers the interference of the multiple rays that come out from the front surface of the sample, considering normal incidence of the light. Two interfaces were considered: air-PS and PS-silicon. The reflection coefficient for multiple reflections can be expressed as:

$$r = \frac{r_1 + r_2 e^{-i\delta}}{1 + r_1 r_2 e^{-i\delta}} = \frac{r_1 + r_2 e^{-i 4\pi n_1 d/\lambda}}{1 + r_1 r_2 e^{-i 4\pi n_1 d/\lambda}} \qquad (1)$$

And the reflectance R is given by:

$$R = r\tilde{r} = \frac{r_1^2 + r_2^2 + 2 r_1 r_2 \cos\delta}{1 + r_1^2 r_2^2 + 2 r_1 r_2 \cos\delta} \qquad (2)$$

In these equations $r_1$ and $r_2$ are the reflection coefficient corresponding to the interfaces air- PS and PS- silicon; $\delta = 4\pi n_1 d/\lambda$ is the phase difference introduced when it crosses the film in the two directions; $n_1$ is the refraction index of the porous silicon and d is the thickness of the layer.

2. Two - layer model. Transference matrix method

Three interfaces were considered: air-ZnTe (I), ZnTe-PS (II) and SP-silicon (III). The boundary conditions of the electric $(\vec{E})$ and magnetic $(\vec{H})$ fields establish that their tangential components have to be continuous through the boundaries. Expressing these conditions in matrix form is possible to obtain the relation between the tangential components of the field in the frontier I and frontier III:

$$\begin{bmatrix} E_I \\ H_I \end{bmatrix} = \mathbb{M} \begin{bmatrix} E_{III} \\ H_{III} \end{bmatrix} \qquad (3)$$

The matrix ($\mathbb{M}$) elements are:

$$m_{11} = Cos[\delta_1] Cos[\delta_2] - Sin[\delta_1] Sin[\delta_2](Y_2/Y_1)$$

$$m_{12} = i(Cos[\delta_2]Sin[\delta_1]/Y_1 + Cos[\delta_1]Sin[\delta_1]/Y_2)$$

$$m_{21} = i(Cos[\delta_2]Sin[\delta_1]Y_1 + Cos[\delta_1]Sin[\delta_1]Y_2)$$

$$m_{22} = Cos[\delta_1]Cos[\delta_2] - Sin[\delta_1]Sin[\delta_2](Y_1/Y_2)$$

Where: $\delta_1 = 2\pi n_1 d_1/\lambda$, $\delta_2 = 2\pi n_2 d_2/\lambda$, $Y_1 = \sqrt{\varepsilon_0/\mu_0}\, n_1$ y $Y_2 = \sqrt{\varepsilon_0/\mu_0}\, n_2$.

$n_1, n_2, d_1$ and $d_2$ are the refractive index and the thickness of the ZnTe and PS layers, respectively. $\varepsilon_0$ and $\mu_0$ are the permittivity and the permeability of vacuum.

Taking into account this expression is possible obtain the reflection coefficient:

$$r = \frac{Y_0 m_{11} + Y_0 Y_s m_{12} - m_{21} - Y_s m_{22}}{Y_0 m_{11} + Y_0 Y_s m_{12} + m_{21} + Y_s m_{22}} \tag{4}$$

Here $Y_0 = \sqrt{\varepsilon_0/\mu_0}\, n_0$ and $Y_s = \sqrt{\varepsilon_0/\mu_0}\, n_s$ while $n_0$ and $n_s$ are the refractive index of the air and the silicon substrate, respectively. Then the reflectivity is calculated as $r\tilde{r}$.

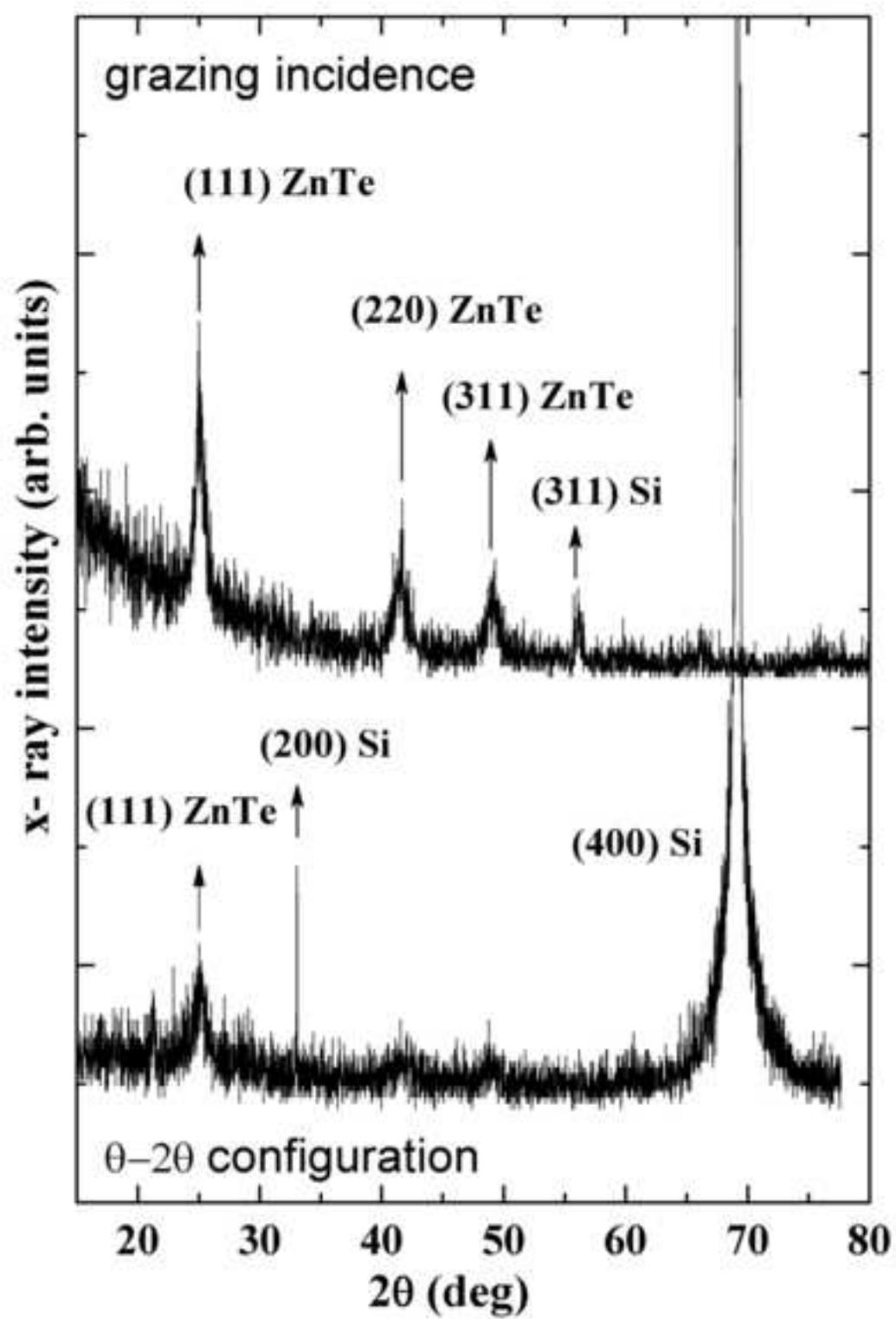

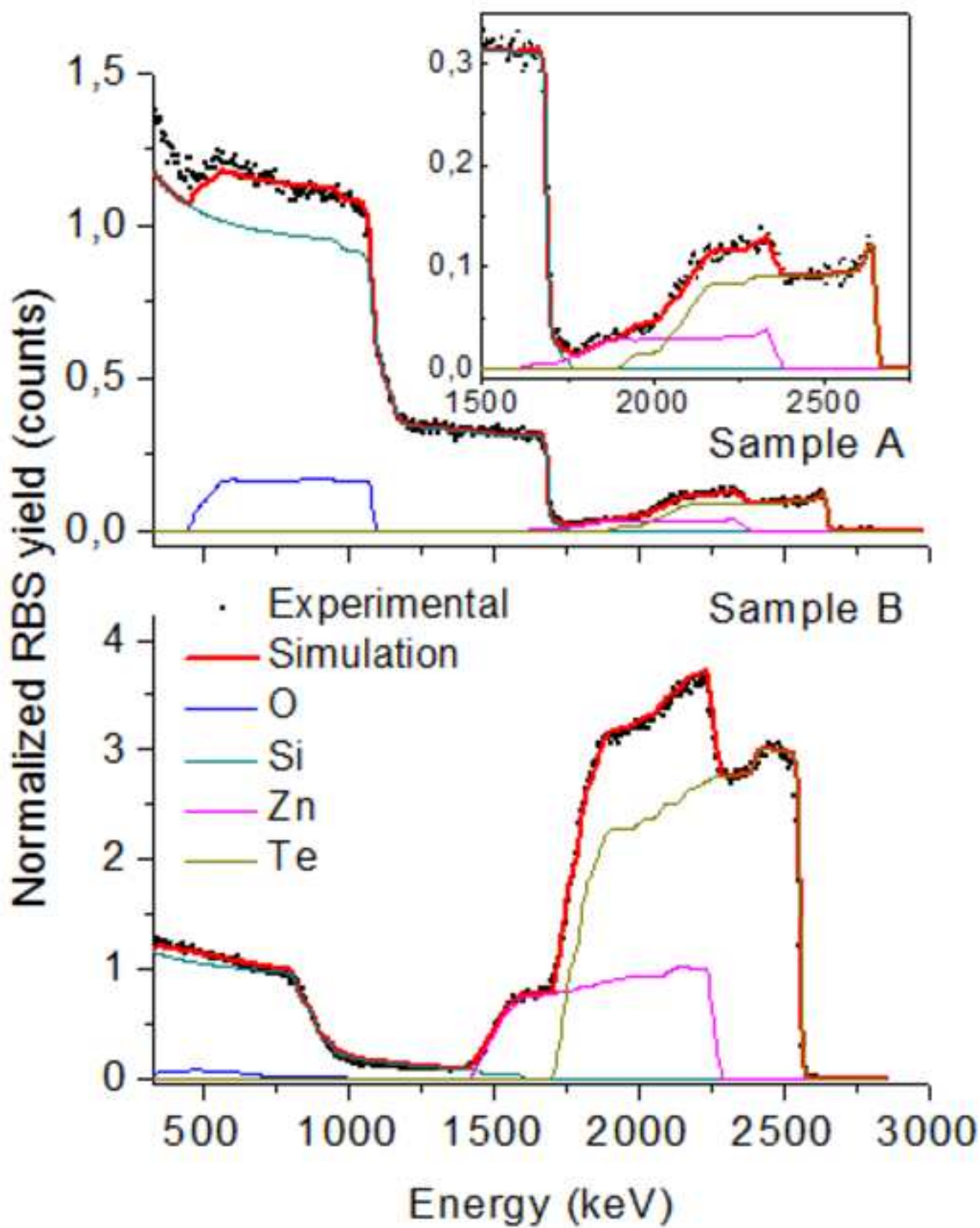

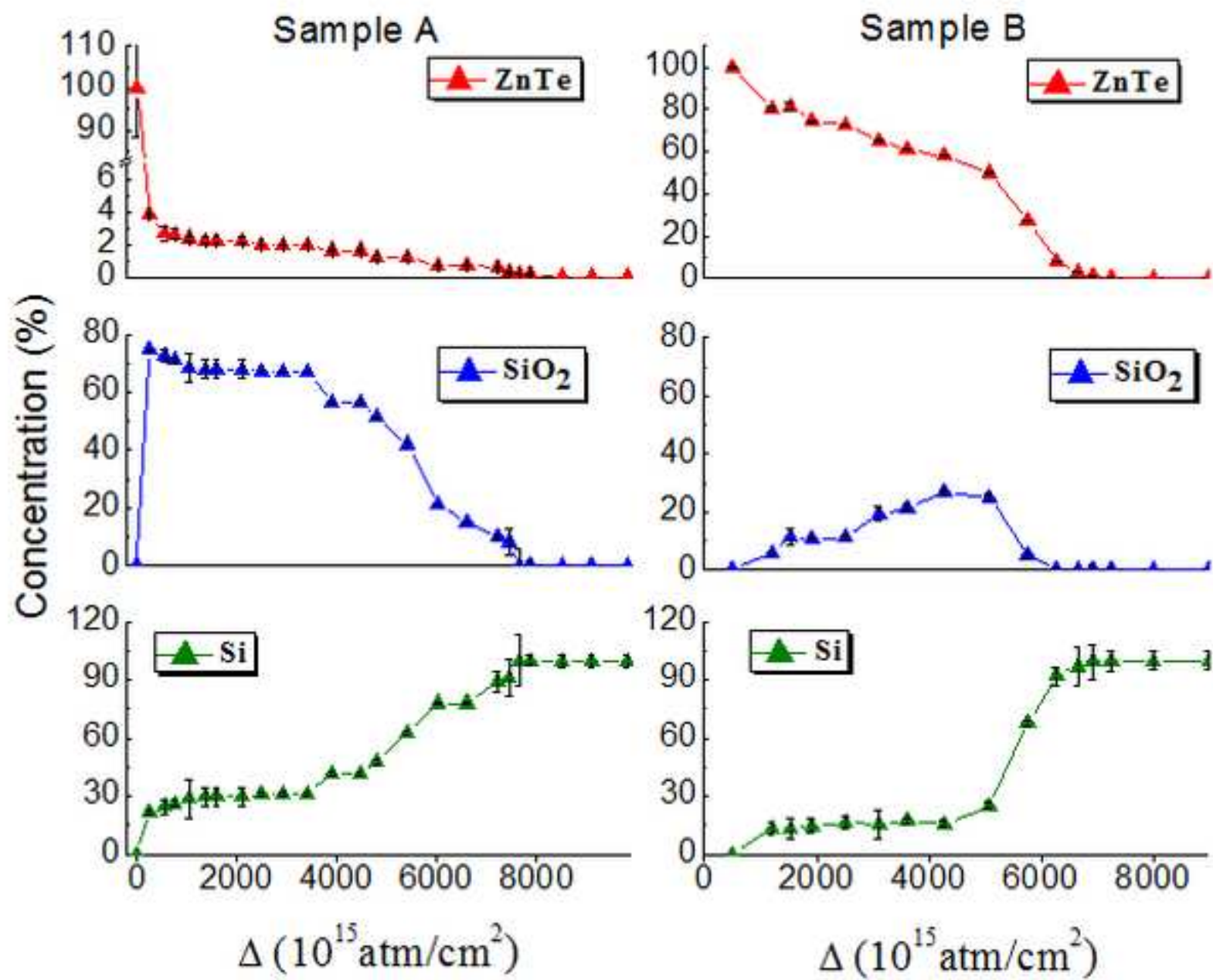

**Fig. 4**

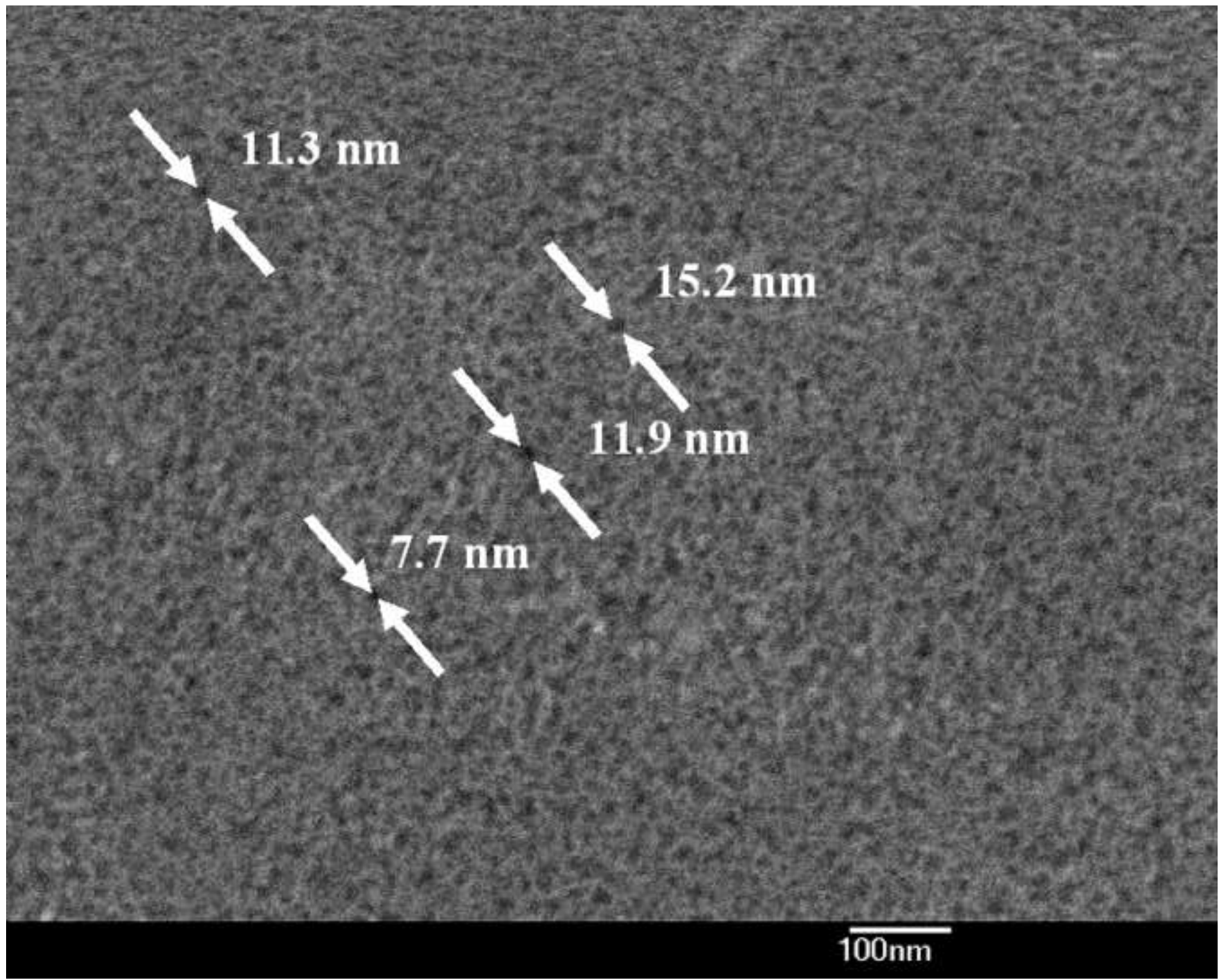

**Fig. 5**
[Click here to download high resolution image](#)

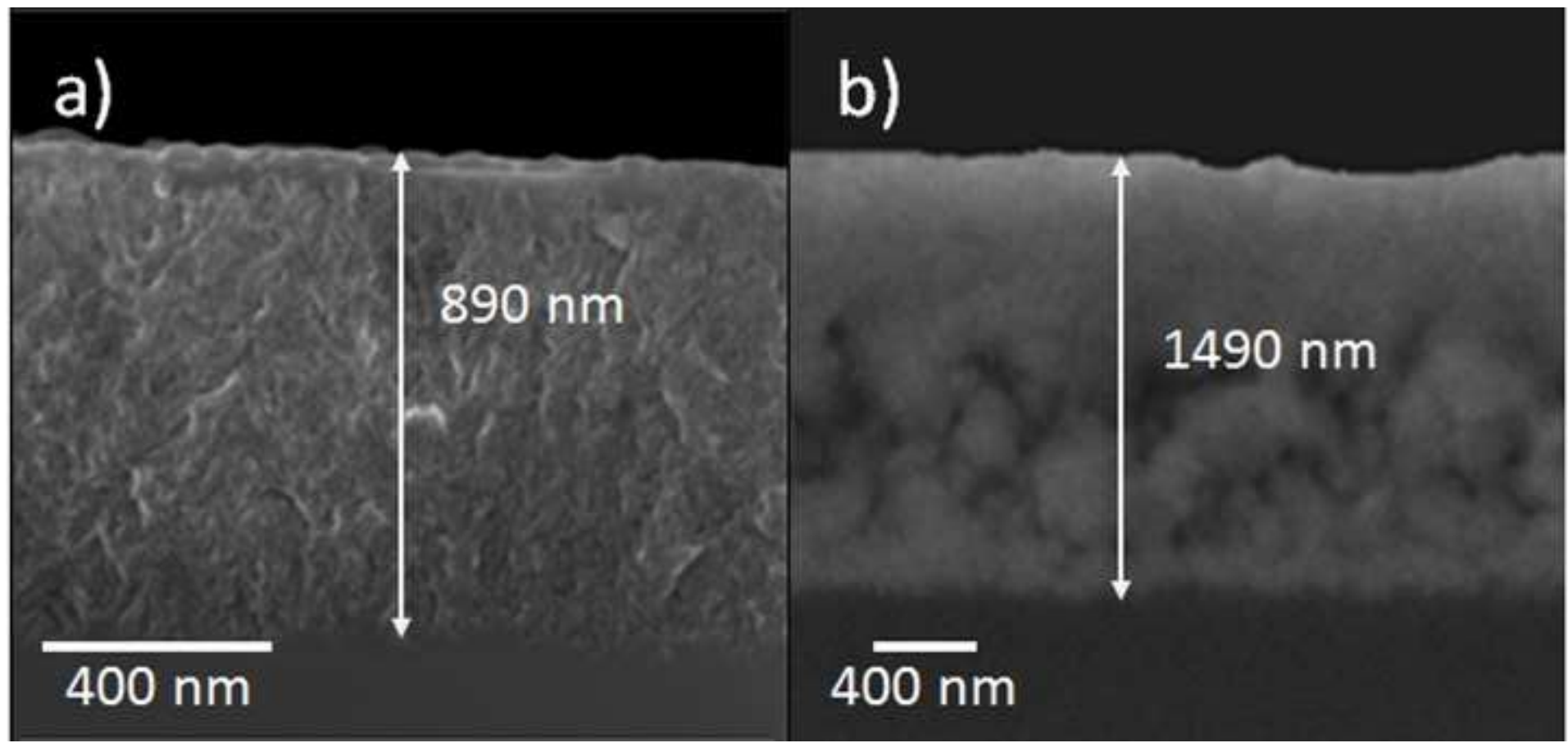

**Fig. 6**
[Click here to download high resolution image](#)

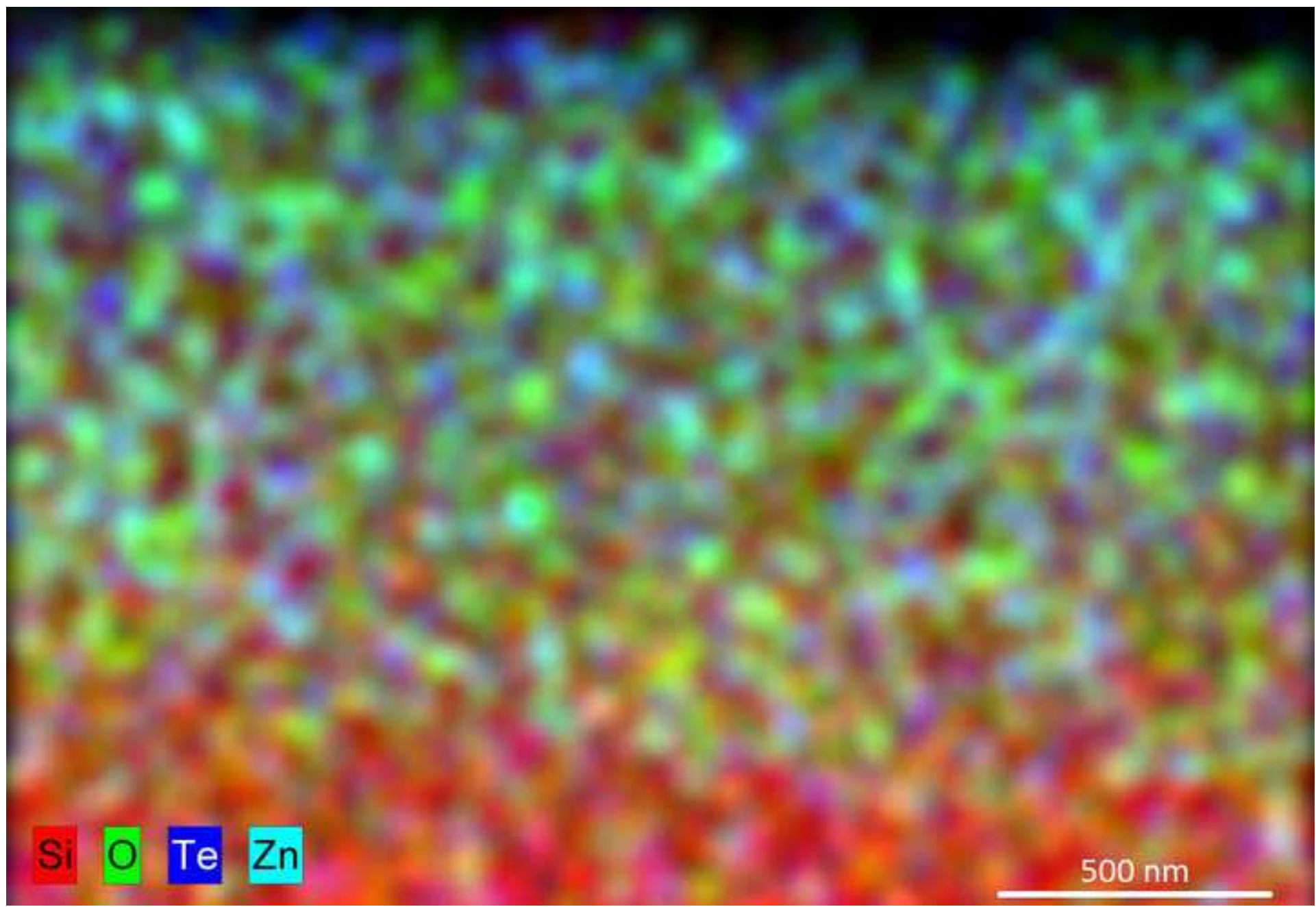

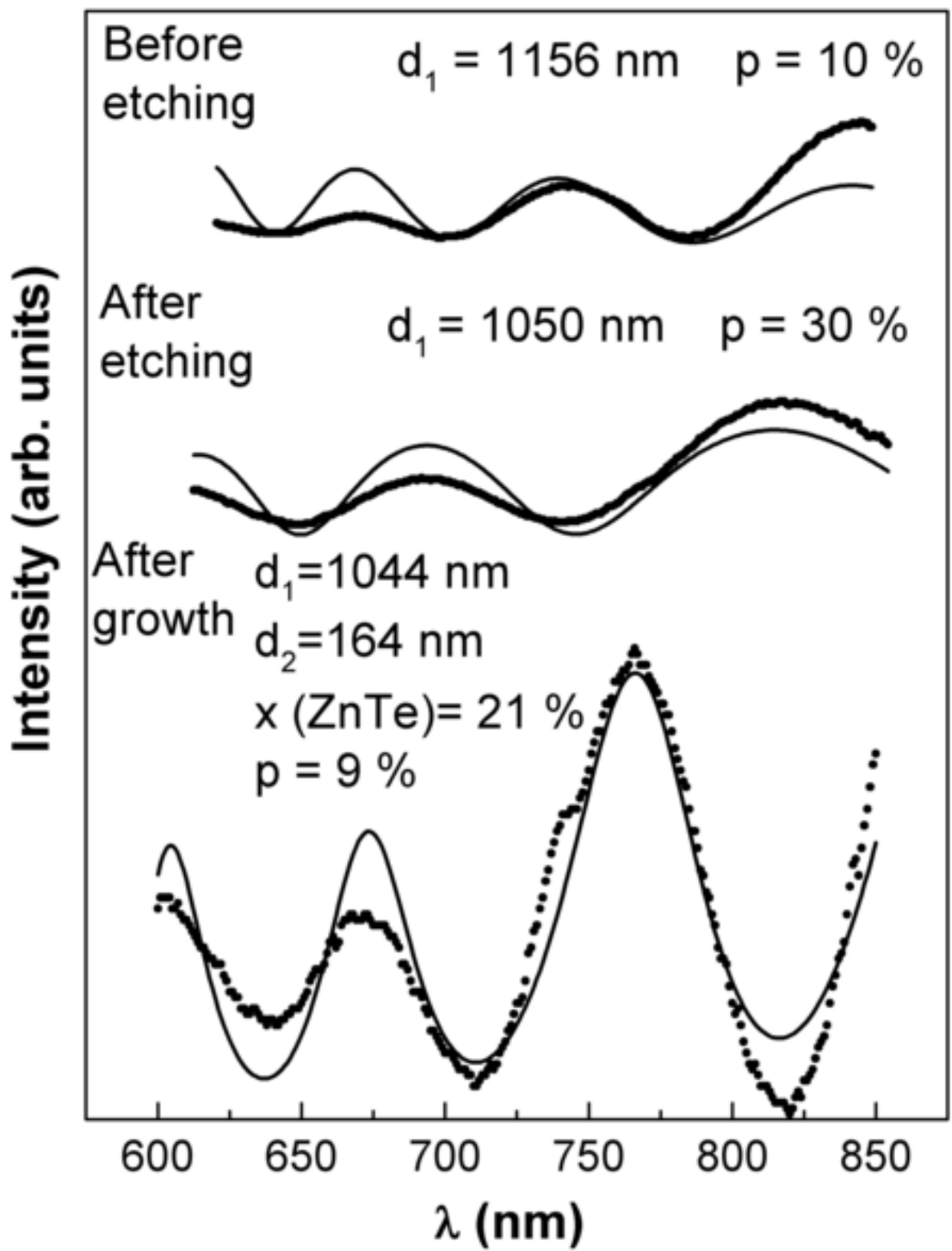